\documentclass[11pt,pra,preprint,superscriptaddress,showpacs,floatfix]{revtex4}

\usepackage{graphicx}
\usepackage{amssymb}
\usepackage{times}
\usepackage{amsmath}
\usepackage{amsthm}
\usepackage{natbib}

\input epsf.tex
\usepackage{graphicx}
\usepackage{amsthm}

\begin{document}

\title{Pseudo antibunching on one single photon detector}

\author{Jianhong Shi\footnote{*Email: purewater@sjtu.edu.cn
		},Yuxing Liu, Guihua Zeng}\affiliation
 {State Key Laboratory of Advanced Optical Communication Systems and Networks, Shanghai Key Laboratory on Navigation and Location-based Service and Center of Quantum Information Sensing and Processing, Shanghai Jiao Tong University, Shanghai, P.R.China, 200240
 }

\date{\today}

\begin{abstract}

 In this paper, we proposed a pseudo antibunching effect on one single photon detector. Though this pseudo antibunching effect is not a sign of the non classical properties of the light field as the antibunching effect. It will give some intresting properties to the photon statistics of the related fields, such as the anticorrelation of photon number or intensity fluctuations of the two input fields. This effect may have some potential application in future especially in the quantum information field.

\end{abstract}

\maketitle

\section{Introduction}\label{Introduction}

It is well known that a single photon source(SPS) exhibits photon antibunching effect; that is,
a dead time between successive photon emission events [1]. This effect can be demonstrated by Hanbury-Brown and Twiss (HBT) type photon correlation measurements.
A HBT experiment setup \cite {Brown1956}, usually consist of one photon source under measurement which is splitted into two arms, with each of them photo-detected individually by single photon detector(SPD) as schematically shown in Fig 1(a). For a pulsed laser excited SPS, the absent of the peak at zero time dalay of  the measured second order temporal correlation function indicates a single-photon emitter.


Opposite to the HBT experiment setup with one photon source and two single photon detectors, we may also consider a reversed HBT setup containing two photon sources and one single photon detector as shown schematically in Fig 1 (b).
A single photon detector also exists a well known 'dead time', which is defined as the amount of time required for the detector to return to its initial quiescent state after a photon event is detected. Thus, for two synchonized pulsed sources with certain delay, a deadtime between successive photon detection event may also possible to cause the absent of the peak at zero time dalay of the measured second order temporal cross correlation function. Typical single photon detector can be described as click detector, which will give an output response(a click) if it detects one or more photons, but can't distinguish two or more photons from one photon.
Then this effect is possible to occur between any type of pulsed sources as soon as there are synchronized. Because this effect is not a direct sign of the non classicality of the light field, we term this as pseudo antibunching effect.

Some interesting properties of this effect are discussed analytically and demonstrated experimentally. We believe this effect will have potential applications in the future.

\begin{figure}[!ht]
	\centering	
	\includegraphics[width=0.8\textwidth]{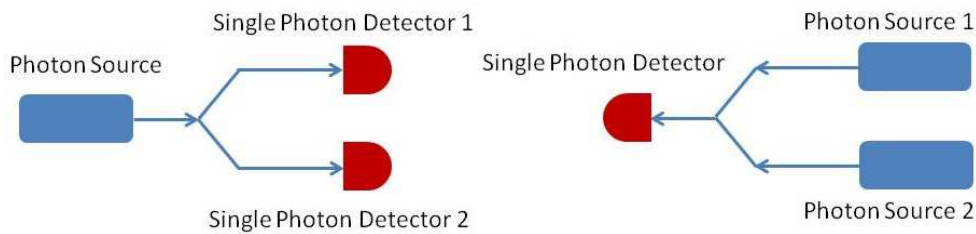}
		\caption{Schematic diagram of HBT setup with one photon source and two single photon detectors (a) and reversed HBT setup with two photon sources and one single photon detector (b)}\label{Fig.1}
\end{figure}

This paper is organized as follows. In Sec. II, we will give the basic theory of pseudo antibunching effect on single photon detector. Simulation and experiment results of the properties of this effect are given in Sec. III. Discussions and conclutions are given in Sec. IV.

\section{Theory}\label{Theory}
 The second order correlation $ g^{(2)} $
  which is also referred to as second order coherence is a common parameter for characterizing the statistics properties of light sources.
The generalized second order correlation between beam i, measured at position $ \vec{r_{1}} $  and time $ t_{1} $, and beam j, measured at position  $ \vec{r_{2}} $ and time $ t_{2} $ can be expressed as[3]
\begin{equation}\label{Eq1}
g_{ij}^{(2)}(\vec{r_{1}},t_{1}; \vec{r_{2}},t_{2})=\dfrac{\langle\vec{E}^{(-)}_{i}(\vec{r_{1}},t_{1})\vec{E}^{(-)}_{j}(\vec{r_{2}},t_{2})\vec{E}^{(+)}_{j}(\vec{r_{2}},t_{2})\vec{E}^{(+)}_{i}(\vec{r_{1}},t_{1})}{\langle\vec{E}^{(-)}_{i}(\vec{r_{1}},t_{1})\vec{E}^{(+)}_{j}(\vec{r_{1}},t_{1})\rangle\langle\vec{E}^{(-)}_{j}(\vec{r_{2}},t_{2})\vec{E}^{(+)}_{j}(\vec{r_{2}},t_{2})\rangle}
\end{equation}
Where $ \vec{E}^{(-)} $  and $ \vec{E}^{(+)} $  correspond to negative and positive-frequency field operators of the detection events and the angled brackets denotes an average over time, while $ i = j $ , $ g^{(2)} $ is the intrabeam second order correlation, while $ i\neq j $, $ g^{(2)} $  denotes the interbeam second order correlation.

For stationary sources, we can write Eq.(\ref{Eq1}) in terms of creation and annihilation operators, and this becomes
\begin{equation}\label{Eq2}
g_{ij}^{(2)}( \vec{r_{1}},t_{1}; \vec{r_{2}},t_{2})=\dfrac{\langle\hat{a}^{\dag}_{i}(\vec{r_{1}},t_{1})\hat{a}^{\dag}_{j}(\vec{r_{2}},t_{2})\hat{a}_{j}(\vec{r_{2}},t_{2})\hat{a}_{i}(\vec{r_{1}},t_{1})\rangle}{\langle\hat{a}^{\dag}_{i}(\vec{r_{1}},t_{1})\hat{a}_{i}(\vec{r_{1}},t_{1})\rangle\langle\hat{a}^{\dag}_{j}(\vec{r_{2}},t_{2})\hat{a}_{j}(\vec{r_{2}},t_{2})\rangle}
\end{equation}

And for one stationary source, measured at one single position or two positions that can be regarded as equivalent such as the measurement in the HBT experiment, $ g^{(2)} $  just denotes the second order temporal correlation:
\begin{equation}\label{Eq3}
g^{(2)}(\tau)=\dfrac{\langle\hat{a}_{i}^{\dag}(t)\hat{a}_{j}^{\dag}(t+\tau)\hat{a}_{j}(t+\tau)\hat{a}_{i}(t)\rangle}{\langle\hat{a}_{i}^{\dag}(t)\hat{a}_{i}(t)\rangle\langle\hat{a}_{j}^{\dag}(t+\tau)\hat{a}_{j}(t+\tau)\rangle}
\end{equation}

For two stationary sources, it's impossible to measure the second order temporal correlation at one single position since they can not be discriminated from each other. But for two  synchronized
pulsed sources, it's possible to measure the second order temporal correlation between them with one click detector if we are able to tell them apart. For two  synchronized pulsed sources which are inherently not stationary, Eq. (\ref{Eq3}) is not valid, It’s convenient to define a discrete version of second order temporal correlation[4,5].
\begin{equation}\label{Eq4}
g_{ij}^{(2)}[k]=\dfrac{\langle\hat{a}_{i}^{\dag}[m]\hat{a}_{j}^{\dag}[m+k]\hat{a}_{j}[m+k]\hat{a}_{i}[m]\rangle}{\langle\hat{a}_{i}^{\dag}[m]\hat{a}_{i}^{\dag}[m]\rangle\langle\hat{a}_{j}^{\dag}[m+k]\hat{a}_{j}^{\dag}[m+k]\rangle}
\end{equation}

Where m and k are integers denoting the pulse number. Square brackets are used to distinguish the discrete form $ g_{ij}^{(2)}[k] $, valid for pulsed sources, from the continuous form $ g_{ij}^{(2)}(\tau) $, valid for CW sources.

In HBT setup with two single photon detectors, we can define the measured discrete temporal correlation function at zero delay for pulsed source as \cite{Steven2013}
\begin{equation}\label{Eq5}
g_{click}^{(2)}[0]=\dfrac{p_{12}(click,click)}{p_{1}(click)p_{2}(click)}
\end{equation}

 Where  $ p_{12}(click,click) $ is the probability that both detector 1 and 2 clicks during the same pulse; $ p_{_1}(click) $ is the probability that detector 1 clicks during a pulse, independent of whether detector 2 clicks; and $ p_{_2}(click) $ is the probability that detector 2 clicks during a pulse, independent of whether detector 1 clicks.The subscipt click used here indicating that this is a property measured by single photon detectors.

An ideal pulsed single photon source can be demonstrated by HBT setup with the measured quantity $ g_{click}^{(2)}[0]=0 $ since one single photon is impossible to trigger two detectors simultaneously.


 In a reversed HBT setup with two sources and one click detector, we can still use Eq.(5) to define the measured discrete temporal correlation function at zero delay for two two  synchronized
 pulsed sources except that the subscript 1,2 stands for the clicks of source 1 and 2 instead of detector 1 and 2. Since we use merely one single photon detector, the information can be used to discriminate the sources of the detection click is the photon arriving time only.

 It's possible to measure the intrabeam second order correlation of one pulsed light source with merely one single photon detector by split this source to two arms and make the time delay between them longer than the dead time of the detector \cite {Dixon2009}. For two independent synchronized pulsed sources, the real second order temporal correlation at zero time delay should always be 1 according to the definition. But from Eq.(5), while the time delay between these two sources is
 shorter than the dead time of the single photon detector, the measured quantity $ g_{click}^{(2)}[0] $ will always be 0 because the single photon detector can only be
 trigger once during one dead time period. In this case, an absent of the peak at zero time delay of the second order correlation function, which is similar to the photon bunching effect of a single photon source will occur.

 Anyway, this $ g_{click}^{(2)}[0] =0 $ phenomena is just a measured quantity no matter what type are these two sources. It can not be treated as the sign of the non classicality of the light field as photon antibunching effect obeserved in HBT setup. So we term this effect pseudo antibunching effect.

Since second order correlation can be used to characterizing the statistics properties of the light source, this pseudo antibunching effect of the measured quantity of $g^{(2)}$ will also affect the measured statistics properties of the light sources.

Now let's consider we have two  synchronized pulsed coherent sources. The photon number distribution of pulsed coherent sources follow a Poisson probability distribution
\begin{equation}
P_i(n) = \dfrac{e^{-\mu_i}\mu_i^n}{n!}
\end{equation}

Where $\mu_i$ are the mean photon number per pulse of the source respectively and the following equation satisfied
\begin{equation}
\mu_{i}=\sum^{\infty}_{0} np_{i}(n)
\end{equation}

A real single photon detector of efficiency $\eta$ can be considered as a perfect detector of unitary efficiency followed with a beam splitter of transmittance $\eta$. While we measure those sources independently by a single photon detector, the measured mean photon number per pulse $\mu_{i,(\mbox{Measure independently})}$ should be the following if we ignore the dark counts and after pulsing of the detector and the pulse period T is larger than the detector dead time.
\begin{equation}
\mu_{i,(\mbox{Measure independently})} =1-e^{-\mu_i\eta_i}
\end{equation}

Since $\mu_i$ are independent with each other, the measured quantity will independent with each other too.

Now let's measure those sources simultaneously with a reversed HBT setup. Suppose the period of the synchronized pulsed sources is T, the time delay between these sources is $T_{delay}$ and the dead time of the detector is $T_{dead}$. For simplicity, assume the width of both pulses is smaller compared with
$T_{dead}$, $T_{delay}$ and $T$ and take one pulse as a whole. In case while $T_{dead}<T_{delay}<<T $, the measured quantity will remain the same since it is equvilent to measure them independently by two detectors. But in the case while $T_{delay}<T_{dead}<<T $, pseudo antibunching effect occurs. Then the measured mean photon number per pulse for the first and the second pulse will be the following
\begin{eqnarray}
\nonumber & \mu_{1,(\mbox{Measure simultanously})}   =  1-e^{-\mu_1\eta_1}  = \mu_{1,(\mbox{Measure independently})}
 \\ \nonumber & \mu_{2,(\mbox{Measure simultanously})}  = e^{-\mu_1\eta_1}(1-e^{-\mu_2\eta_2})
\\  &=(1-\mu_{1,(\mbox{Measure independently})})\mu_{2,(\mbox{Measure independently})}
\end{eqnarray}

From Eq. (9), we can see that though the real statistics properties of these two sources are independent, the measured properties will be related. And in the above situation we discussed, the measured mean photon number of the first pulse remains the same as we measure those pulses independently. But the measured mean photon number of the second pulse will have an anticorrelation with the measured mean photon number of the first pulse. For example, in a extrem case while $\mu_{1}$ is very big and $\mu_{1,(\mbox{Measure independently})}$ is almost 1, $\mu_{2,(\mbox{Measure simultanously})} $ will close to 0 even if $\mu_{2}(\mbox{Measure independently})$ is big.

\section{Simulation and experiment}\label{simulation}

In following simulation and experiment, pulsed coherent sources are used. Suppose two synchronized pulsed coherent sources are measured in a reversed HBT setup with one single photon detector. For simplicity, we assume that $T_{delay}<T_{dead}<<T $ and a $100\% $ detection efficiency of the single photon detector.
Assume that the real mean photon number per pulse(pulse intensity) of pulse 2 is proportional to pulse 1 and increase simultanously.
Figure.2 shows the real mean photon number per pulse of pulse 1 Vs the measured mean photon number per pulse of pulse 2 with 50\%-200\% portion. While measure pulse 2 independently with pulse 1 blocked, an exponential relation between the real and measured quantity is expected. While measure them simutanously, the measured quantity of pulse 2 will change according to Eq.(9). In this case, the measured quantity will not increase monotonically with the real quantity. Fig.3 shows the measured mean photon number per pulse of pulse 1 Vs pulse 2.
\begin{figure}[!h]
	\centering
	
	\includegraphics[width=0.7\textwidth]{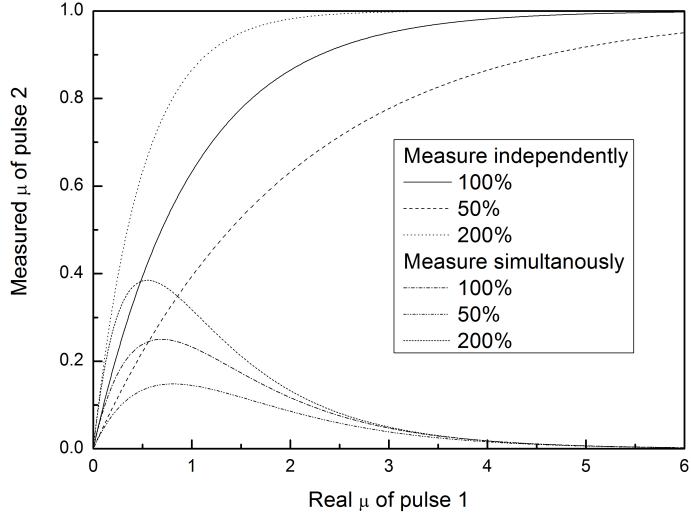}
	\caption{Real Vs. Measured mean photon number per pulse in reversed HBT setup for two pulsed coherent sources with the intensity of the pulses increase simultanously}\label{Fig.2}
\end{figure}

\begin{figure}[!h]
	\centering
	
	\includegraphics[width=0.7\textwidth]{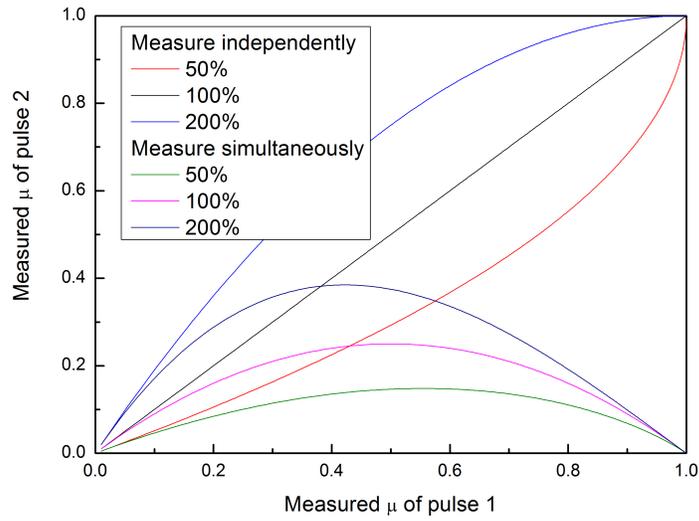}
	\caption{Measured mean photon number per pulse of pulse 1 vs Measured mean photon number per pulse of pulse 2}\label{Fig.3}
\end{figure}

\begin{figure}[!h]
	\centering
	
	\includegraphics[width=0.8\textwidth]{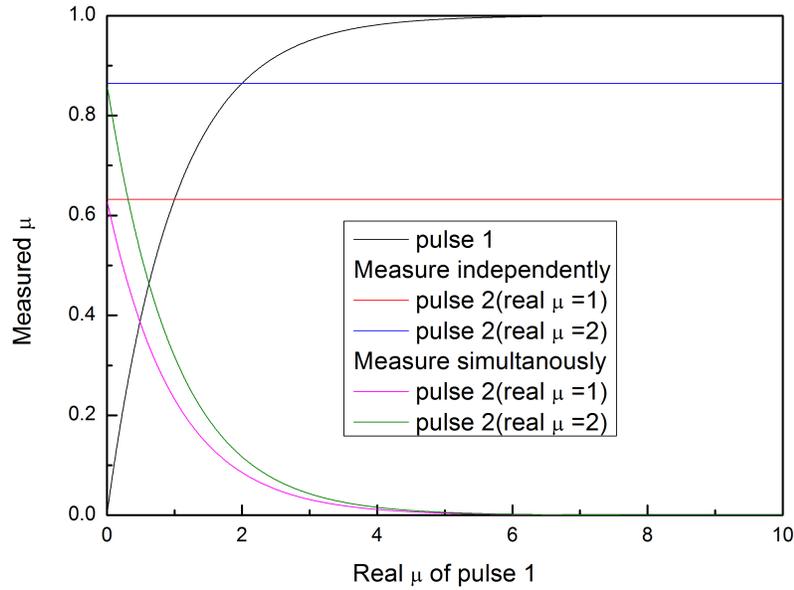}/
	\caption{Real Vs. Measured mean photon number per pulse with reversed HBT setup while intensity of pulse 2 keep stable}\label{Fig.4}
\end{figure}

In Figure 4 and 5, we keep the intensity of pulse 2 unchanged. While measure it independently, the measured mean photon number per pulse will keep stable. While measure those two pulses simultanously, a clear anticorrelation between them caused by pseudo antibuching effect is shown in Fig 5 as indicate by Eq.(9). In fig.4 and 5, the real mean photon number per pulse of pulse 2 is set to be 1(2) and the measured quantity without pulse 1 is about 0.62(0.86).

\begin{figure}[!h]
	\centering
	
	\includegraphics[width=0.8\textwidth]{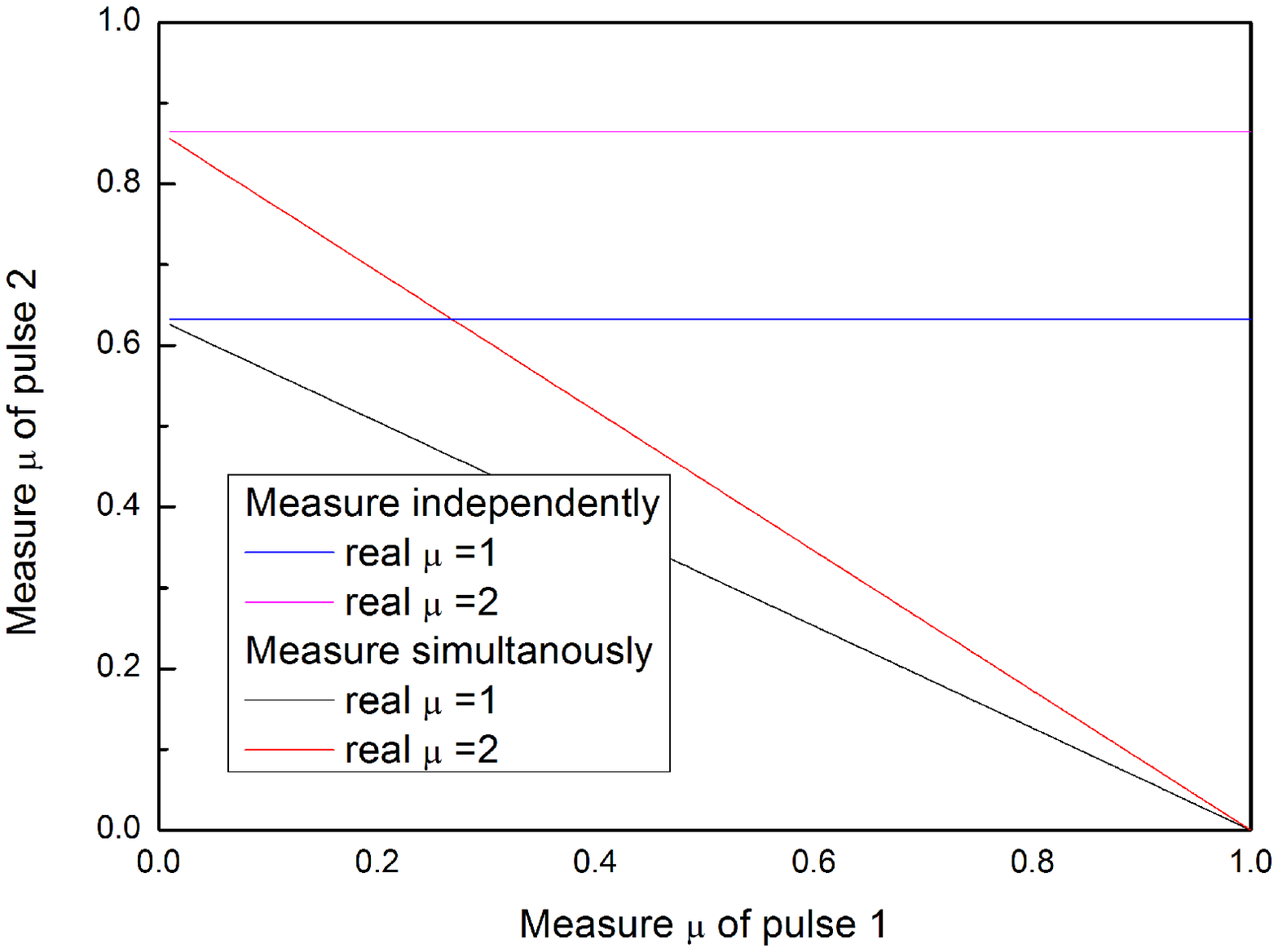}
	\caption{Measured mean photon number per pulse of pulse 1 Vs. pulse 2 while intensity of pulse 2 keep stable}\label{Fig.5}
\end{figure}

\begin{figure}[!h]
	\centering
	
	\includegraphics[width=0.8\textwidth]{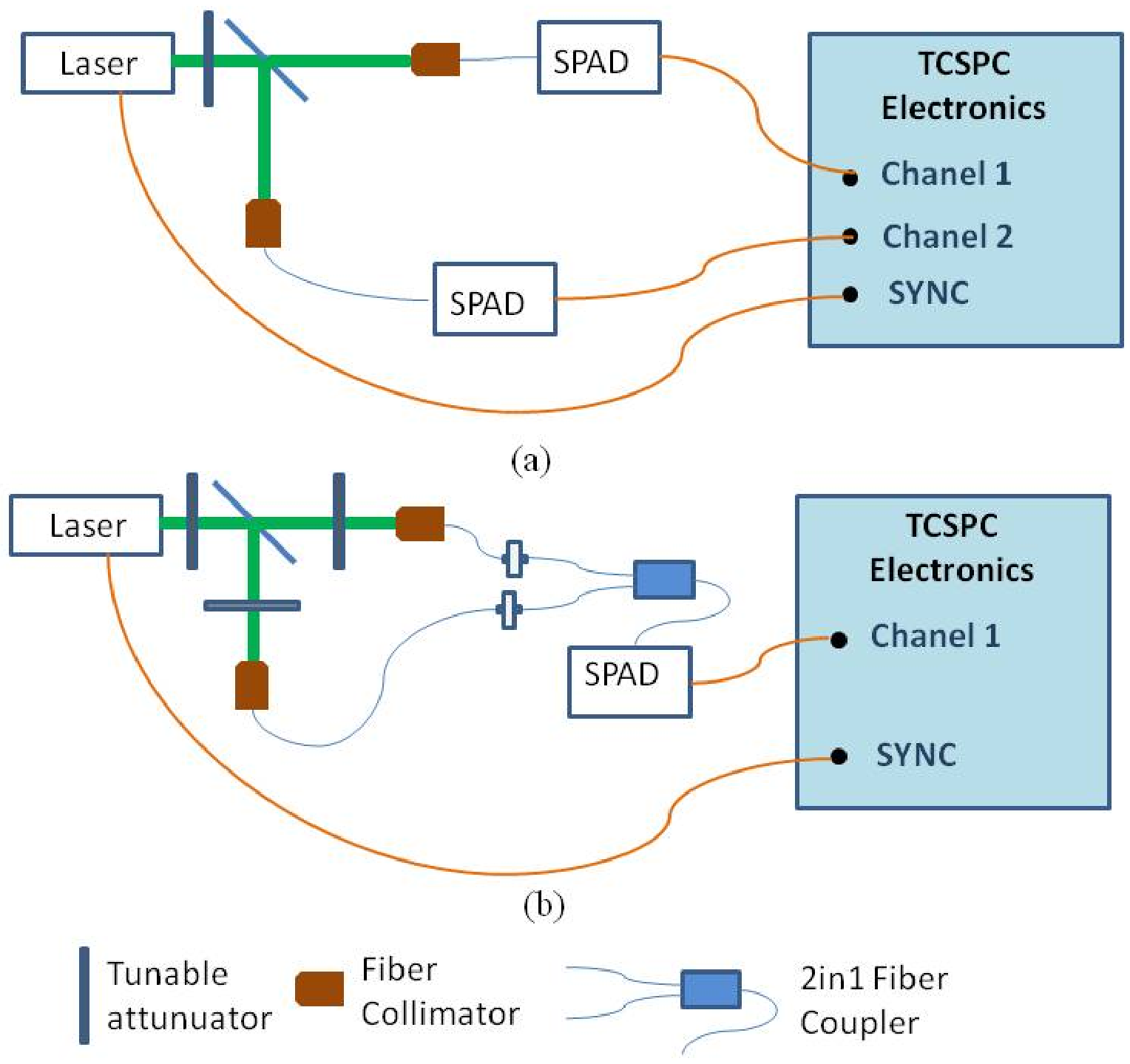}/
	\caption{Experimental setup}\label{Fig.6}
\end{figure}
Figure 6 shows the experimental setup. One attenuated pulsed source is split into two arms by a 50:50 non-polarized beam splitter. In Fig 6(a), two arms are detected by two single photon avalanche detector(SPAD) individually, which is just the HBT type second order correlation measurement setup. In Fig 6(b), two arms are combined together with fiber coupler and detected by one SPAD with certain delay introduced between them, which is an adjusted reversed HBT setup. Tunable attenuators are used before and after the beam splitter to attunuate the intensity of the pulses in both arms simultanously and independently.
Output of each SPAD is connected to one input channel of the TCSPC module of Pico Quant HydraHarp 400. The electrical synchronized signal from the laser is connected to the "sync in" channel of HydraHarp 400. A Time-Tagged Time-Resolved(TTTR) mode is used to measure the data. With this mode, we can record the arriving time of every detected photon event originating from multiple detectors or multiple sources with respect to both the start of the experiment and to a common excitation sync pulse. Then we can perform a post software based analysis, such as to calculate the second order temporal correlation function at any delay time or get the measured mean photon number per pulse of those pulses.

In our experiment, pulsed supercontinum laser source with about 100ps pulse duration is used.The frequency of the laser pulse is 1MHz, which correspond to 1$\mu$s pulse period. The dead time of the single photon detector is 77ns. The delay between two arms is about 5ns, which is introduced
by 1m length multimode fiber. The situation $T_{delay}<T_{dead}<<T $ is satisfied.

\begin{figure}[!h]
	\centering
	
	\includegraphics[width=0.8\textwidth]{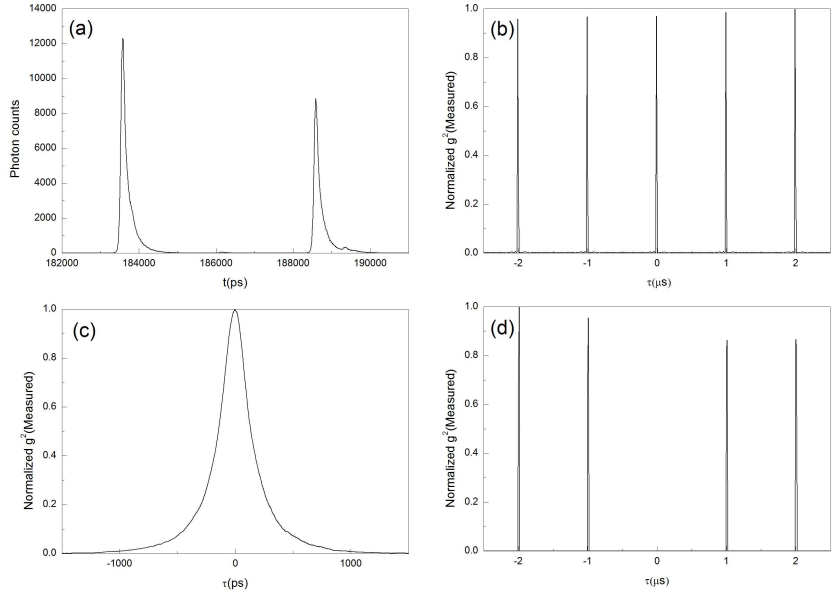}
	\caption{(a)Measure pulse intensity waveform of pulse 1 and 2 (b)measured normalized $g^{(2)}$ for HBT Experiment setup(c)Measured Normalized $g^{(2)}$ for HBT Experiment setup in one period(d)Measured Normalized $g^{(2)}$ in reversed HBT Experiment while measure two arms simultaneously}\label{Fig.7}
\end{figure}

In each experiment, data acquisition time is 1s, which correspond to 1 million pulse periods.
And the intensity of the pulses are carefully attenuated. For HBT setup, owing to the pulsed nature of the source, a comb-like second order correlation function with period 1$\mu$s which is same as the laser source is shown in Fig 7(b). Fig 7(c) shows the correlation function with shorter time gate and finner delay around 0 time delay. For the reversed HBT setup, since $T_{delay} $ is bigger than the pulse duration, clearly seperated peak of two pulses are shown in Fig 7(a), The measured pulse width is about 150ps and the measured pulse delay is about 5ns. Then the originate of the photon detection event can be clearly discriminated. We can sort the photon detection events to each arm and calculate the cross correlation function based on the measured data.
An absent of the peak at zero time delay in the measured second order cross correlation function is shown in Fig 7(d), which demonstrate the pseudo antibunching effect on one single photon detector.

Futhermore, with the reversed HBT setup shown in Fig 6(b), we change the intensity of both pulses simultanously by tuning the attunuator before the beam splitter. In this case,the intensity of the two arms will increase or decrease simultanously with a constant portion. Then we measure these pulses independently with one arm blocked and measure them simultanously.

\begin{table}[htb]
	\centering\caption{Measured photon counts of first pulse and second pulse with reversed HBT setup}
	\begin{tabular}{lp{1.3in}|lp{1.3in}}
		\hline  Measured photon & counts for first pulse & Measured photon  & counts for second pulse \\
		\hline
	Independently & Simultanously & Independently & Simultanously  \\
	 \hline
	     27775 & 27958 & 18534 & 17698  \\
	     64853 & 65059 & 37134 & 34572  \\
		216301 & 221807 & 146464 & 114401 \\
		365105 & 365920 & 262309 & 165292 \\
		403902 & 408171 & 285329 & 169511 \\
		463866 & 468631 & 335863 & 176741 \\
		504933 & 508772 & 368659 & 179484 \\
		595745 & 596597 & 463821 & 183451 \\
		633118 & 637812 & 485174 & 170717 \\
		697912 & 705293 & 547679 & 156606 \\
		845617 & 862366 & 752693 & 94582 \\
		973508 & 969942 & 951434 & 3012 \\		
			
		 \hline
	\end{tabular}
\end{table}

Table 1 shows the measured photon counts of first pulse and second pulse, which can be get by just counting the photon counts in the corresponding time duration of the first pulse and second pulse in 1s sampling time. The measured mean photon number per pulse for each pulse can be get by averaging the photon counts over 1 million pulse periods. We can see the measured photon counts for each pulse increases with the increase of the pulse intensity while we measure them independently with one of the arms blocked. But while we measure these two pulses simultanously, the measured photon counts for the second pulse will change according to Eq.(9) because of the pseudo antibunching effect.

\begin{figure}[!h]
	\centering
	
	\includegraphics[width=0.9\textwidth]{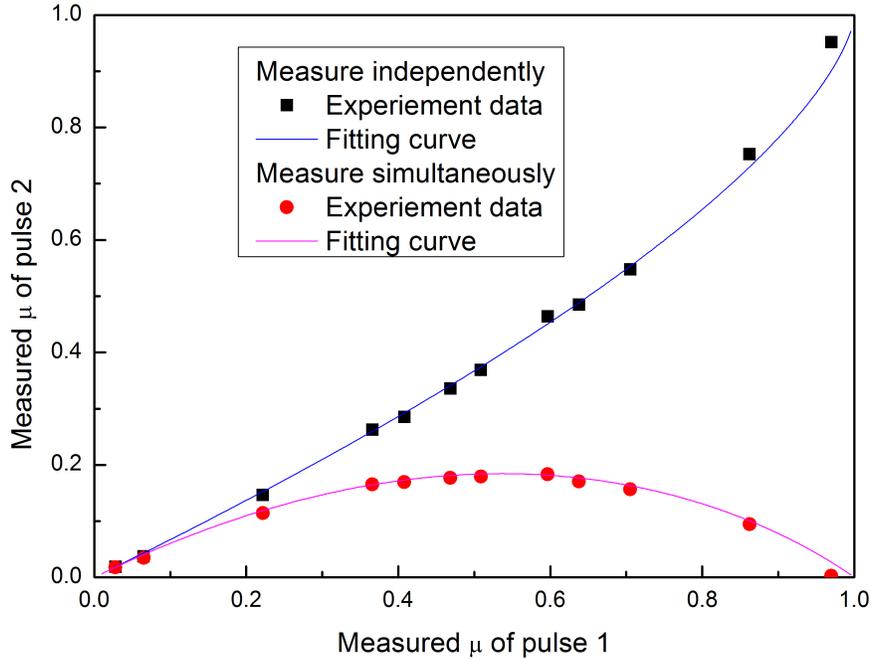}
	\caption{Measured mean photon number per pulse of pulse 1 vs pulse 2 by measure them independently and simultanously}\label{Fig.8}
\end{figure}

Figure 8 shows the measured mean photon number per pulse of pulse 1 Vs. pulse 2 by independent and simultanous measurement. Because of the unblance of the two arms, an intensity portion of 66\% between pulse 2 and pulse 1 cause by the fiber delay line and the coupling efficiency difference of two arms is observed and fit well with the experimental data.

Then we keep the intensity of pulse 2 unchanged and tune the intensity of pulse 1. Fig 9 shows the measured mean photon number per pulse of pulse 1 Vs. pulse 2. A clear anticorrelation between them is observed.

\begin{figure}[!h]
	\centering
	
	\includegraphics[width=0.9\textwidth]{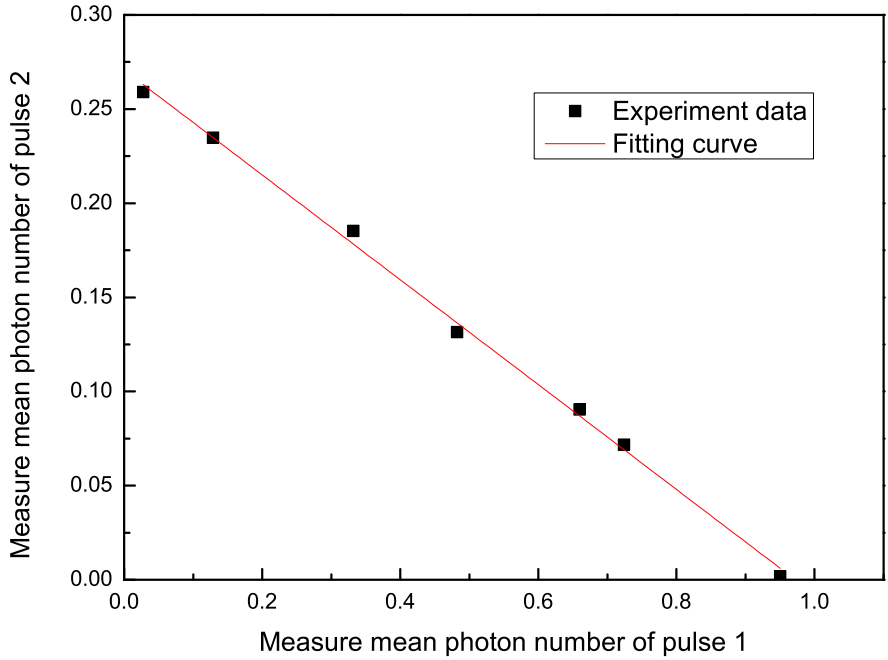}
	\caption{Measured mean photon number of pulse 1 vs pulse 2 with increase intensity of pulse 1}
	\label{Fig.9}
\end{figure}
\section{Discussion and Conclusion}
From the simulation and experiment results, we can see that while two synchronized pulsed sources are measured with one single photon detector simultaneously, the measured quantities will not independent because of the pseudo antibunching effect. An anticorrelation is observed between them while $T_{delay}<T_{dead}<<T $. In this case, for example, the intensity fluctuation of one pulse could be represented with the measured quantity of another synchronized pulse though these pulses are not related before. If any information is contained in the intensity fluctuation of the first pulse, this information could be get by just concern about the measured results of the second pulse.

As we indicated in above content, this pseudo antibunching effect will occur whenever $T_{delay}<T_{dead}$ between two synchronized pulses regardness what properties of those pulses. This effect could also exist between three or more synchronized pulses when we measure them simultanously with one single photon detector. Since this effect is occur on one single photon detector, useful information can be get only if the originate of the sources can be discriminate by photon arriving time.

In conclution, we proposed a pseudo antibunching effect on one single photon detector in this paper. Interesting properties of this effect is theoretically analized and
experimentally demontrated. Anyway, though this pseudo antibunching effect could not be taken as the sign of the non classical light field as the photon antibunching effect, we believe this effect could have some potential application in quantum infomation or some other field in the future.

\subsection{Acknowledgement}

This work was supported by the National Natural Science Foundation of China (Grants No: 61170228, 61332019, 61471239), the Hi-Tech Research and Development Program of China (Grant No: 2013AA122901)



\end{document}